# An Energy Aware Clustering Scheme for 5G-enabled Edge Computing based IoMT Framework

Jitendra Kumar Samriya [1][0000-0001-5466-8899], Mohit Kumar [1][0000-0003-1600-6872],
Maria Ganzha[2][0000-0001-7714-4844], Marcin Paprzycki[2][0000-0002-8069-2152],
Marek Bolanowski[3][0000-0003-4645-967X], Andrzej Paszkiewicz[3][0000-0001-7573-3856]

[1] Dr.B.R.Ambedkar National Institute of Technology, Jalandhar, India.
jitu.samriya@gmail.com kumarmohit@nitj.ac.in
[2] Polish Academy of Sciences, Warszawa, Poland
[3] Rzeszow University of Technology, Rzeszów, Poland

**Abstract.** In recent years, 5G network systems start to offer communication infrastructure for Internet of Things (IoT) applications, especially for health care service providers. In smart health care systems, edge computing enabled Internet of Medical Things (IoMT) is an innovative technology to provide online health care monitoring facility to patients. Here, energy consumption, along with extending the lifespan of biosensor network, is a key concern. In this contribution, a Chicken Swarm Optimization algorithm, based on Energy Efficient Multi-objective clustering scheme is applied in the context of IoMT system. An effective fitness function is designed for cluster head selection., using multiple objectives, such as residual energy, queuing delay, communication cost, link quality and node centrality. Simulated outcomes of the proposed scheme are compared with the existing schemes in terms of parameters such as cluster formation time, energy consumption, network lifetime, throughput and propagation delay.

## 1 Introduction

Within the context of rapid growth of Internet of Things (IoT), availability of 5G networks empowers the epoch of Internet of Everything [1,2]. Along with development of wearable sensor devices, innovative IoT applications and facilities materialize in healthcare domain, being jointly referred to as Internet of Medical Things (IoMT) [3,4]. Typically, the IoMT systems are linked with wireless body area networks (WBAN), in which the sensor network is used to sense the health-related parameters of the human body. In healthcare (IoMT) applications, sensor nodes are often called biosensor nodes [5]. Here, biosensors monitor, among others, heartbeat, electrocardiogram, blood pressure, etc. To function, the IoMT ecosystem needs network connectivity between sensors and control device(s). Note that biosensor often acts like a personal digital assistant (PDA; [6, 7]). They are low power devices, and thus are well suited to use in the vicinity of the human body, as they do not cause negative impact to the human. However, if the energy in the biosensor is exhausted, the entire WBAN will collapse, which results in obvious severe problem [8, 9]. Note also that replacement of the biosensor is rather difficult, when it is placed inside the patient [10]. In this context note that that the goal of energy-efficient clustering protocols is to, among others, achieve effective cluster head selection [11, 12]. However, recent energy aware clustering and routing schemes suffers from network overhead [13, 14]. Separately, fuzzy control based energy efficient clustering protocol [15] enhanced the reliability of the network, but lacks in the energy consumption rate. Moreover, heterogeneity based energy aware clustering protocol have been designed in [16, 17], for IoT based healthcare applications. Even though there are many techniques to make the biosensors energy efficient, but IoMT systems are impacted by the need to deliver stable/secure/flexible solutions. Thus, delivering a robust clustering protocol is required for medical healthcare applications.

The key contribution of this work is to present a clustering approach for IoMT platforms, which offers energy-aware communication in 5G enabled edge-based ecosystems. Particularly, high energy-efficiency is to be obtained during data transmission among wearable sensor nodes. Here, IoMT environment comprises of resource-limited wearable sensor nodes (SNs) which collect the sensitive patient data and direct them towards healthcare service provider, through 5G enabled base station (BS) for further processing. The transmission and reception of data takes more power, so it is crucial to maximize lifespan of network and its energy-sustaining abilities. Hence, multi objective based cluster head (CH) selection process is performed, using Chicken Swarm Optimization (CSO) algorithm, for effective cluster formation. Typical CH selection methods focus on distance, energy, and delay. However, since the proposed clustering is designed for edge computing based IoMT environments, two additional factors are also considered, i.e.: packet queue and node capacity. Therefore, the selection of CH, in the proposed clustering scheme, relies on as residual energy, queuing delay, communication cost, link quality and node centrality of the SNs.

The remainder of this contribution is structured as follows. Related work is summarized in Section 2. Section 3 provides details of the proposed approach. Section 4 illustrates the performance of the proposed scheme. Section 5 concludes the work.

## 2 Related works

Let us start by summarizing recent studies, related to energy-efficient clustering, explored in IoMT ecosystems. Bharathi et.al. ([18]) proposed an energy-efficient clustering, using disease diagnosis design for health care IoT. It was also called Energy Efficient Particle Swarm Optimization based Clustering (EEPSOC). Initially, deployed IoT devices utilized for healthcare data sensing, to be grouped into clusters. Then, CH was selected using EEPSOC. Nasri et.al. ([19]) presented a novel hierarchical energy efficiency routing protocol, in order to improve the energy efficiency of the wireless sensor node, and to enhance the network lifetime. The presented protocol employs fuzzy logic is to optimize wireless sensor networks performance. The proposed cross-layer routing protocol uses data from different layers to select the optimal CH. For efficient routing and network lifespan extension, Majumdar et.al. ([20]) proposed a novel extremal optimization tuned micro genetic (EO-GA) clustering approach, for efficient routing and network lifespan extension, by conserving the rechargeable battery of network edge devices. For a WBAN-based IoT system, Jiang and Li ([21]) suggested an energy-aware sensor node clustering technique. Here, the body region is separated into 3-segments, used in the suggested clustering: lower, upper, and middle. Within every region, an improved LEACH clustering technique is used. The whale optimization algorithm (WOA), is then utilized to intelligently pick cluster heads. For IoMT systems, Sajedi et.al. ([22]) proposed F-LEACH (Fuzzified LEACH) based data aggregation strategy. The F-LEACH aiming at maximizing the network lifetime. For data aggregation, this work uses the dynamic behaviour of the LEACH protocol. Sood et.al. ([23]) developed an energy-efficient clustering approach by applying artificial bee colony (ABC) scheme. The near-optimal CH were selected depending on Euclidean distance among each device and its neighbours, residual energy, Euclidean distance between devices and the sink, and the number of neighbours. Saba et.al. ([24]) presented a secure energy-efficient scheme by IoMT. The energy consumption and communication overhead, among the biosensors, were decreased while sending the healthcare data. First, the IoMT devices were linked in the way of a complete graph, such that there should be a unique edge between each pair of nodes. Then, the sub-graphs were extracted by Kruskal's algorithm by calculating the minimum cost. The routing decision was optimized from IoMT sensors to medical centers. Han et.al. ([25]) developed a clustering design for medical applications (CMMA). Here, for CH election, to give an effective communication for IoMT, CMMA considers devices queue and capacity, along with distance, energy, and delay for the CH selection procedure. A clustering based swarm Intelligence approach was introduced by El-Shafeiy et.al. ([26]) for periodically managing, analyzing, clustering, and discovering useful information about potential patients in IoMT systems (calle, SIoMT). The SIoMT approach deals with the cluster nodes hosting and processes of optimization. Bee Colony Optimization algorithm (BCO) scheme automatically grouped the clusters by modifying the main parameters of the network. Anguraj et.al. ([27]) have presented an Enriched CH selection, by augmented bifold cuckoo search algorithm (ABCSA), for edge-based IoMT. The solutions in the given search space was handled and operated by novel binary model. The effectiveness of the CH selection was improved by newly proposed fitness function. The binary solutions were handled by changing to best choices, rather than using continuous values. The uniformity was achieved by the CH selection along with low energy consumption of nodes. Table 1 presents a summary of the algorithms discussed thus far.

Table 1. Comparison of energy-efficient clustering algorithms explored in medical IoT

| Ref. | Aim | Parameters | Future Scope |
|---|---|---|---|
| [18] | To obtain energy efficient cluster head | energy and distance | To incorporate a compressive sensing technique |
| [19] | To enhance the energy level and the system lifetime | residual energy and node distance | To propose solution for security enhancement |
| [20] | To enhance the lifetime of the edge network | energy, sleeping time, distance to CH over data analytics center, computational load, and intra-cluster distance | Energy minimization |
| [21] | To maximize network lifespan and save network energy through reducing sending and receiving duplicate packets | residual energy and node distance | To improve energy efficiency and node distance |
| [22] | To maximize network lifetime | distance and energy | - |
| [23] | To minimize utilization of energy and delay of transmission | energy and distance | To propose an delay-aware and energy-efficient routing scheme |
| [24] | To minimize communication overhead and energy consumption | distance and energy | To validate proposed model via mobility-based medical scenarios |
| [25] | To facilitate active communication for IoMT-based applications | energy, delay, distance, capacity and packet queue | To improve energy efficient and joint secure background for remote health observing |
| [26] | To perform network optimization | node distance | To explored on other IoT datasets |
| [27] | To enhance network lifetime | distance and energy | To propose joint secure energy-based framework |

## 3   System model and assumptions

The deployed wearable sensors are assumed to monitor human physical activities and collect health-related data, and forward it to stationary BS. Additionally, it is assumed that uniform level of energy is allocated to all wearable sensor nodes and energy needed for a node to perform intra-cluster communication is represented by any arbitrary value within the pre-determined range. Moreover, sleeping mode is considered, when there's no activity happening in the monitoring environment of IoMT ecosystem. The diminution in lifespan of network happens when sensor node battery is being drained. Hence, the energy-efficiency problem is taken into account to elect the CH, amid the accessible SNs. The model of the system illustrating the proposed clustering scheme is depicted in Fig. 1.

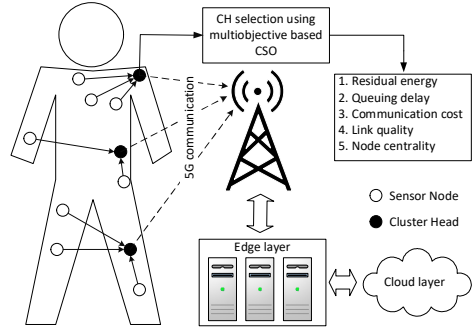

**Fig. 1.** Proposed Clustering Scheme System Model Framework

In this paper, the energy model discussed in [28], has been selected. It is assumes that each IoMT network deployed wearable sensor node utilizes energy during data forwarding and receiving process. The equation for calculating energy consumption of data packet of size $s$ bits for overall distance($d$) is $E_{Trans}(d) = (TA_{FS}d^\alpha + E_D)s$. $E_D$ denotes the energy consumption of an electronic device, $TA_{FS}$ signifies the free space model amplifier of a transmitter, and $\alpha$ denotes the path loss exponents, with $2 \leq \alpha \leq 4$. Energy utilization of receiver to obtain the data packet is represented as $E_{Rec}(d) = s \times E_D$. The cumulative energy utilization of each wearable sensor node, supposed to send or receive data, is based on distance $d$ is represented as $E_{Cum} = \{(TA)d^\alpha + 2(E_D)\}s$. The selection of cluster head primarily relies on the objective function values, which are minimal or maximal. In this work, the selection of energy efficient CH depends on multiple objectives, namely residual energy, queuing delay, communication cost, link quality and node centrality.

**Residual Energy:** Initially, the wearable SNs, deployed inside the IoMT network, gather the sensitive patient's data and forward them to the relevant CH. The role of the CH is to obtain the data from SNs and then sent them to the BS. The energy consumption of CHs, during the period of data gathering from SNs, can be computed as: $E_{CH-SN} = D_B \times \left(E_{PB_F} + AE_{PB} \times \left(\sqrt{(a_{CH} - a_{SN})^2 + (b_{CH} - b_{SN})^2}\right)\right)$, where $(a_{CH}, b_{CH})$ is the position of CH and $(a_{SN}, b_{SN})$ is the position of SN, $D_B$ is the total number of bits in the data packet, $E_{PB_F}$ is the energy needed, per bit, for data forwarding, and $AE_{PB}$ is the amplification energy, needed per bit for data forwarding. The data forwarding between CH to BS can be computed as follows: $E_{BS-} = D_B \times \left(E_{PB_F}\left(\frac{N}{Y} - 1\right) + \left(E_{PB_G} \times \left(\frac{N}{Y}\right)\right) + E_{PB_F} + AE_{PB} \times \left(\sqrt{(a_{BS} - a_{CH})^2 + (b_{BS} - b_{CH})^2}\right)\right)$, where $(a_{BS} - b_{BS})$ is the position of BS, $E_{PB_F}$ is the energy needed, per bit, for data forwarding, $N$ is the total number of *SN*s in the IoMT system, $Y$ denotes the number of SNs in the cluster. Finally, the cumulative energy consumption of each cluster is computed as: $E_C = E_{BS-C} + \left(\left(\frac{N}{Y}\right) - 1\right) \times E_{CH-SN}$.

**Communication Cost:** The commination cost is defined as the power utilized for data forwarding. It is directly proportional to square of the distance between the source and the sink nodes. The evaluation of communication cost performed as follows: $Com_C = \frac{d_{avg}^2}{d_0^2}$, where $d_{avg}$ denotes the average distance between given node and neighbor nodes and $d_0$ represents the forwarding radius of a node.

**Queuing Delay:** Another important parameter for cluster head selection is queuing delay $D_{Que}$, which depends on the packets arrival rate to the SN, and the outward link forwarding capacity. Assume that $A_R$ is the arrival rate of packets $P_i$ to SN and $F_C$ is forwarding capacity of outer link. Then, queuing delay $D_{Que}$ can be calculated as: $D_{Que} = (A_R + F_C)/P_i$.

**Link Quality:** In IoMT system, the fading of channel is irregular and time-varying. If the receiver does not receive the signal precisely, re-forwarding will happen. Re-forwarding require additional energy from the transmitter. Therefore, the link quality is computed for attaining fair energy efficiency. It is estimated by equation: $LQ = \frac{LQ_i - LQ_{min}}{LQ_{max} - LQ_{min}}$, where $LQ_{max}$ and $LQ_{min}$ denotes the upper and lower range of re-forwarding from neighbors, respectively; and $LQ_i$ represents entire re-transmission cost among neighbors and given node.

**Node centrality:** The node centrality measure $i$ determines the number of times a node behaves as a link on the shortest paths among two other nodes. It is computed as: $N_C = \sum_{m \neq r \neq n \in R} \frac{\lambda_{mn(i)}}{\lambda_{mn}}$, where $\lambda_{mn}$ is the cumulative number of shortest paths between node $m$ and $n$, and $\lambda_{mn(i)}$ is the number of paths via $i$. The outcome values range from 0 to 1. At last, every considered node follows the fitness function based on calculated objective function values, along with the weighted coefficients to limit the final cost value for CH selection: $Fitness_{final} = w_1 \times E_C + w_2 \times \left(\frac{1}{Com_C}\right) + w_3 \times \left(\frac{1}{D_{Que}}\right) + w_4 \times LQ + w_5 \times N_C$.

Generally, $w_1 + w_2 + w_3 + w_4 + w_5 = 1$ and, $0 \leq w_i \leq 1, \forall i, 1 \leq i \leq 5$. Here, the central goal is to maximize the fitness given as follows: $Maximize \sum_{i=1}^{|CH|} Fitness_{final}$ such that $1 \leq i \leq |CH|$. The considered node, which fulfills all objectives will be selected as a CH i.e., the node with maximum residual energy, link quality and node centrality at minimum queuing delay, and communication cost. The selected CH in every cluster is accountable for data gathering and forwarding data packets to BS directly or via inter-cluster communication. After the selection of CH in every cluster, the route will be discovered for transferring the collected data to BS.

### 3.1 Chicken swarm optimization (CSO) algorithm for CH selection

To optimize parameters of residual energy, queuing delay, communication cost, link quality and node centrality, for CH selection, the CSO algorithm is proposed. The proposed model is based on the algorithm proposed in [29]. The most important aspects of CSO, in the context of solving the CH selection problem, are as follows.

***Chickens' Movement:*** best node is assigned as the rooster and the worst node as the chick, while the remaining nodes are hens. Let $R_n$ indicates the count of roosters, $H_n$ indicates the count of hens, $C_n$ indicates the count of chicks, and $M_n$ indicates the count of mother hens in the swarm of chickens, while $B$ – is the number of iterations. The chickens' positions can be identified using $c_{U_{i,j}}^{t_a}$ where $i \in [1,2,\ldots\ldots N]$ and $j \in [1,2,\ldots\ldots D]$ for a time $t_a$ in $D$ dimensional space. In our work, the simulated rooster can be identified as the CH with optimal fitness value.

***Roosters' Movement:*** Following [29] movement of roosters can be computed as:

$$c_{U_{i,j}}^{t_a+1} = c_{U_{i,j}}^{t_a} \times [1 + Randn(0, \sigma^2)] \tag{1}$$

$$\sigma^2 = \begin{cases} 1 & \text{if } f_i \leq f_k \\ \exp\left(\frac{f_k - f_i}{|f_i| + \varepsilon}\right) & \text{otherwise}; k \in [1, N], k \neq i \end{cases}$$

where $c_{U_{i,j}}^{t_a+1}$ depicts the movement of the rooster, $Randn(0, \sigma^2)$ denotes the Gaussian distribution, with mean value 0 and standard deviation $\sigma^2$, $\varepsilon$ denotes a constant value added in order to avoid error due to zero-division, $k$ implies the index of the rooster selected randomly from the group and $f_i$ denotes the value of fitness of the rooster $x_i$.

***Hens' Movement:*** Following [29] hen movement can be mathematically as:

$$c_{U_{i,j}}^{t_a+1} = c_{U_{i,j}}^{t_a} + S1 \times Rand \times \left(c_{U_{r1,j}}^{t_a} - c_{U_{i,j}}^{t_a}\right) + S2 \times Rand \times \left(c_{U_{r2,j}}^{t_a} - c_{U_{i,j}}^{t_a}\right) \tag{2}$$

where $S1 = \exp\left(\frac{f_i - f_{r1}}{abs(f_i + \varepsilon)}\right)$, $S2 = \exp(f_{r2} - f_i)$ and *Rand* represents a random number in [0, 1], $r1 \in [1,2,\ldots..N]$ indicates the index of the mate of $i^{th}$ hen, $r2 \in [1,2,\ldots\ldots N]$ indicates the index of the rooster or hen chosen randomly from the swarm, $S1$ and $S2$ indicate the influence factors.

***Chicks' Movement:*** Following [29] this aspect can be formulated as:

$$c_{U_{i,j}}^{t_a+1} = c_{U_{i,j}}^{t_a} + FL \times \left(c_{U_{m,j}}^{t_a} - c_{U_{i,j}}^{t_a}\right) \tag{3}$$

where $c_{U_{m,j}}^{t_a}$ denotes the location of the mother of $i^{th}$ chick so that $m \in [1,2,\ldots.N]$, $FL$ denotes the speed of the chick following the mother, the value of $FL$ is taken randomly between 0 and 2 to indicate the differences among chicken.

For selecting the CH among the available sensor nodes, the accessible SNs are considered to be the chicken swarm, where the nodes with best fitness values are chosen to be roosters, and with worst fitness values are chicks, while the remaining nodes are known to be hens. For each exploration of the search space, the location of the rooster is updated using formula (1). Following the rooster, the hens forage and the location of every hen is updated using formula (2). The chicks searching for food around their mother will also explore new search spaces, which is captured using (3). For the established swarm, the ranking of the chickens is performed to maintain hierarchical order. Based on the fitness values computed, chickens are ranked, leading to proper location update for each exploration. After ranking the chickens, the relationship among the mothers and chicks are identified to find out the differences between the chicks. Algorithm 1 depicts the proposed CSO based algorithm for optimal CH selection. The algorithm facilitates the selection of optimal CH to effectively form the clusters to attain energy efficiency as well as long life span of network. In the proposed IoMT based system, after selection of CH, using CSO algorithm, the BS disseminate a beacon message to identify the CH to form a cluster. The wearable sensor node, which satisfies the fitness value of the CSO algorithm will be declared as a CH, while remaining SNs are declared as cluster members. The CH selection results in each SN obtaining position as CH or cluster member. After this process is completed, the TDMA slot of every cluster member node is used to transfer the gathered patient sensitive data to the CH. Moreover, inactive SNs move to the sleep state to sustain energy. At last, after CHs forward data to BSs, data is removed by wearable sensors.

**Algorithm 1: Multi objective based CSO Algorithm for CH selection**
**Input:** N number of CHs, CSO parameters; **Output:** Pareto Solution S indicating the nodes that act as CHs.

1. Initialize all the parameters $R_n, H_n, C_n, M_n$ and $B$
2. Initialize the chickens in the swarm randomly as $C_{U_i}(i = 1,2,\ldots\ldots y)$
3. Initialize the total count of iterations as $Max_{itr}$
4. While $T_r < Max_{itr}$ do
5. If $(T_r \% B = 0)$ then
6. Establish the hierarchical order through ranking of chickens
7. Partition the swarm group and identify the mother-child relationship
8. End if
9. For $(i = 1)$ do
10. If $(i == rooster)$ do
11. Perform local update of the rooster's location using (1)
12. End if
13. If $(i == hen)$ do
14. Perform local update of the hen's location using (2)
15. End if
16. If $(i == chick)$ do
17. Perform local update of the chick's location using (3)
18. End if
19. Estimate the fitness of the obtained solution using $Fitness_{final}$
20. If the solution outperforms the older one → update location
21. End for
22. Label the best solution as pareto optimal solution S
23. End while
24. Return S

## 4 Experimental results and discussion

This section presents the details of simulation setup and performance of proposed multi objective based clustering scheme. The performance is measured using:
  a) *Cluster formation time:* time taken to form the cluster.
  b) *Energy Consumption:* defines the amount of energy consumed by the SNs. Here, after each round the overall energy consumption (in mJ) is measured.
  c) *Network lifetime:* signifies for how much time, or in how many rounds, the network can accomplish its processes.
  d) *Throughput:* number of bits transferred to BS from CHs (Mb/s).
  e) *Delay:* time taken by the data to be transferred from SN to BS with the aid of CH, i.e., the distance between the SN and the BS divided by the speed of the signal expressed in milliseconds (ms).

The proposed clustering scheme has been simulated using MATLAB 2018a. It was run on a system with the Intel(R) Core(TM) i5 CPU 7500 processor with a clock rate of 3.40 GHZ and 6 GB of main memory. The performance of proposed CSO based clustering scheme was compared with the existing EO-μGA [20], ABCSA [27], BCO [26] and PSO [18] based schemes. The details of simulation and algorithm parameters are listed in Tables 2 and 3 respectively.

**Table 2.** IoMT network parameters

| Parameters | Parameters |
|---|---|
| Total Number of SNs | 1000 |
| IoMT sensing area | $500 * 500\ m^2$ |
| Base Station Position | (500,500) |
| Packets Size | 1500 bits |
| Max Network Throughput | 1 Mbps |
| Initial Node Energy | 2J |
| Electronics energy | $30\ nJ/bit$ |
| Data aggregation energy | $3\ nJ/bit/signal$ |
| Transmitting power | $9\ mW$ |
| Max number of rounds | 500 |

**Table 3.** CSO algorithm parameters

| Parameters | Parameters |
|---|---|
| Population Size | 100 |
| Number of roasters | 3 |
| Number of hens | 5 |
| Update time steps | 10 |
| Maximum Iterations | 150 |

As is shown in Fig. 2, proposed CSO-based clustering scheme has the lowest cluster formation time, while the PSO has the highest cluster formation time. The reason is that the CSO has fast convergence speed, hence forms the clusters effectively with the aid of robust fitness function, which comprises multiple objective function values. The other methods, presented in Fig. 2, have higher cluster formation time, compared to the proposed scheme due to the low convergence speed (related also to their fitness functions).

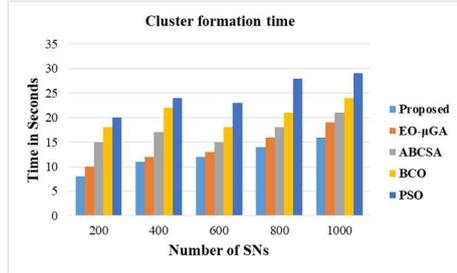

Fig. 2. Cluster Formation Time

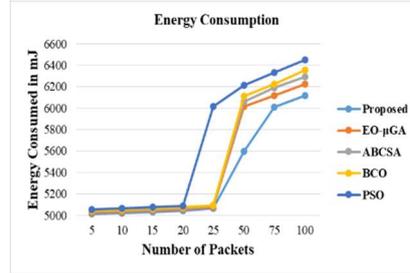

Fig. 3. Energy consumption at varying number of packets

As what concerns energy consumption, simulation based on varying (from 5 to 100) the number of packets was conducted. The performance of energy consumption achieved by each schemes are shown in Fig.3. When the number of packets increases, energy consumption also increases for all schemes. The proposed CSO minimizes the cost by 1.9%, 2.7%, 3.8% and 4.9% in comparison to EO-μGA, ABCSA, BCO and PSO, respectively.

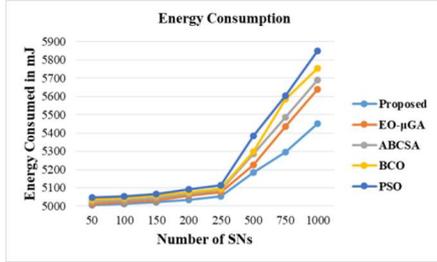 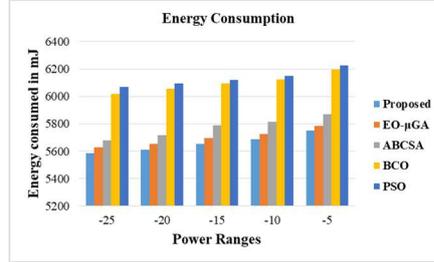

Fig. 4. Energy consumption at varying number of SNs

Fig. 5. Energy consumption at varying the transmission power ranges

Next, the influence of the number of sensors (50 to 1000) was considered form the point of view of energy consumption (see, Fig. 4). Obviously, energy consumption increases when the number of nodes increase. Here, the proposed scheme minimize the consumption of energy by 3.4%, 4.3%, 5.5% and 7.1% in comparison to EO-µGA, ABCSA, BCO and PSO, respectively.

Subsequent series of experiments evaluated energy consumption when the transmission power ranged from $-25dBm$ to $-5dBm$ (see, Fig. 5). Proposed scheme outperformed other schemes for all transmission power values.

To evaluate the network lifetime performance of proposed scheme, we considered for the number of sensor nodes varying from 50 to 1000 (see, Fig. 6).

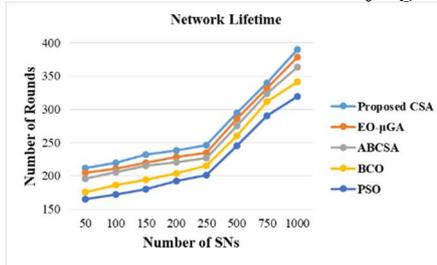 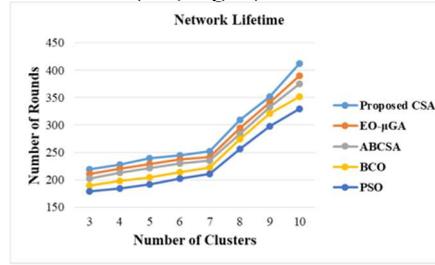

Fig. 6. Network Lifetime at varying number of SNs

Fig. 7. Network Lifetime at varying number of Clusters

The network lifetime of the proposed approach achieved fair improvements over existing schemes. Specifically, it improved network lifetime by 3.2%, 5.4%, 10.4%, and 17% over EO-µGA, ABCSA, BCO and PSO, respectively.

Network lifetime was also evaluated with respect to the number of clusters (from 3 to 10; see, Fig.7). Here, the proposed schemes performs well and maximizes the network lifetime by 5.7%, 10.4%, 16.4% and 21.3% in comparison to EO-µGA, ABCSA, BCO and PSO, respectively.

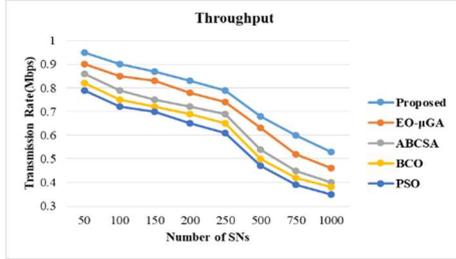 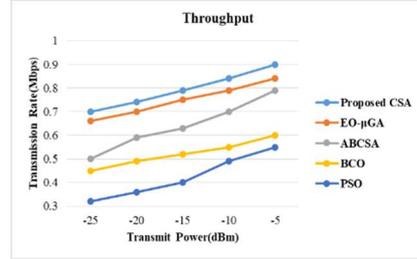

Fig. 8. Throughput at varying number of SNs    Fig. 9. Throughput at varying the transmission power ranges

In the next series of experiments, throughput was simulated when varying the sensor nodes from 50 to 1000 (see, Fig. 8). The proposed scheme provides better throughput performance. Specifically, the performance is higher by 0.1%, 26%.31.1% and 39% over EO-μGA, ABCSA, BCO and PSO, respectively.

Throughput was also evaluated when varying the ranges of transmission power from $-25dBm$ to $-5dBm$ (see, Fig. 9). The proposed schemes performs well and improves the throughput by 6.8%, 13%, 40% and 48,2% in comparison to EO-μGA, ABCSA, BCO and PSO, respectively.

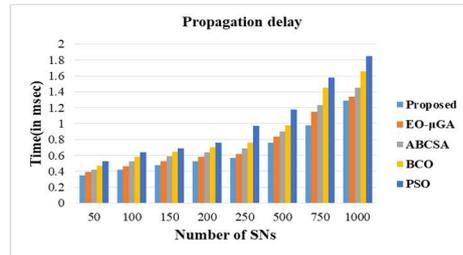

Fig. 10. Propagation delay at varying the number of SNs

Finally, Fig. 10 depicts propagation delay with respect to varying number of sensor nodes, from 50 to 1000. The simulations results confirm that proposed CSO based clustering scheme makes propagation delay 0.05 milliseconds lower than EO-μGA, 0.16 milliseconds lower than ABCSA, 0.35 milliseconds lower than BCO and 0.56 milliseconds lower than PSO.

## 5    Concluding remarks

In this work, an energy efficient CSO-based clustering scheme was proposed for IoMT ecosystems. Since IoMT is installed to frequently gather patients' health related data, conserving energy of each wearable SNs is necessary to enhance the lifespan of the system. The proposed clustering scheme uses an effective fitness function, based on residual energy, queuing delay, communication cost, link quality and node centrality. The sensor node with maximum fitness across multiple objectives is selected as the cluster head, using CSO algorithm. The performance of the proposed CSO-based clustering scheme has been evaluated and compared with the existing EO-μGA, ABCSA, BCO and PSO based approaches, in terms of cluster formation time, energy consumption, network lifetime, throughput and propagation delay. The simulation outcomes shows that the proposed clustering scheme (1) outperformed the remaining approaches in all measured characteristics, and, specifically, (2) brought about 3-7% reduction in energy consumption. In the future, the proposed scheme will be extended with respect to mobility of nodes, body actions, and cross layer optimization for complex network situation.